\documentclass[11pt,a4paper]{article}
\usepackage{amsmath,amssymb,amsfonts,a4wide,graphicx,bm,times}
\usepackage[colorlinks=true,linkcolor=black, citecolor=black,
urlcolor=black]{hyperref} \input{epsf}
\numberwithin{equation}{section}
\makeatletter \let\old@startsection=\@startsection
\renewcommand{\@startsection}[6]
{\old@startsection{#1}{#2}{#3}{#4}{#5}{#6\mathversion{bold}}}
\makeatother

\usepackage[utf8]{inputenc}

\def\Tr{{\rm tr\, }}
\def\defeq{\stackrel{\text{def}}=}
\newcommand\re[1]{({\ref{#1}})}
\def\be{\begin{eqnarray}}
    \def\ee{\end{eqnarray}}
    \def\no{\nonumber}
    \def\la{\label}
    \def\l {\l}
\def\({\left(} \def\){\right)} 
\def\<{\left\langle\,} 
\def\>{\, \right\rangle} 
\def\[{\left[}
 \def\]{\right]} 

    \def\hf{ {\textstyle{1\over 2}} }
    \def\sqrti{ {\textstyle{1\over \sqrt{2}}} }

\def\d{\delta} 

  \def\p{\partial}

  \def\s{\sigma}
  \def\t{\tau}

  \def\CF{{\cal F}}

 \def\O{\Omega} 

 \def\mm {_{_{(+)}}}

\def\CZ{{\cal Z}}

\def\l{z}

\def\CC{ {\cal C}}

\def\p{\partial}

\def\s{\sigma}
\def\IZ{ {\mathbb Z}}

\def\o{ \omega }

\def\M{{\bf M}}
\def\rtwv{ |0_{\text{tw}}\rangle }
\def\ltwv{ \langle 0_{\text{tw}}|}

 \def\mm{ \mu}

\begin{document}

\thispagestyle{empty}

\begin{flushright}
IPhT-t09/150\\
\end{flushright}

\vspace{1cm}
\setcounter{footnote}{0}

\begin{center}

 {\Large\bf Matrix Models as  CFT:    Genus
  Expansion}

\vspace{20mm} 

Ivan Kostov\footnote{\it Associate member of the Institute for Nuclear
Research and Nuclear Energy, Bulgarian Academy of Sciences, 72
Tsarigradsko Chauss\'ee, 1784 Sofia, Bulgaria}\\[7mm]

{\it   Institut de Physique Th\'eorique, CNRS-URA 2306 \\
	     C.E.A.-Saclay, \\
	     F-91191 Gif-sur-Yvette, France \\[5mm]
}
     \end{center}

\vskip9mm

\vskip18mm

 \noindent{ We show how the formulation of the matrix models as
 conformal field theories on a Riemann surfaces can be used to compute
 the genus expansion of the observables.  Here we consider the
 simplest example of the hermitian matrix model, where the classical
 solution is described by a hyperelliptic Riemann surface.  To each
 branch point of the Riemann surface we associate an operator which
 represents a twist field dressed by the modes of the twisted boson.
 The partition function of the matrix model is computed as a
 correlation function of such dressed twist fields.  The perturbative
 construction of the dressing operators yields a set of Feynman rules
 for the genus expansion, which involve vertices, propagators and
 tadpoles.  The vertices are universal, the propagators and the
 tadpoles depend on the the Riemann surface.  As a demonstration we
 evaluate the genus-two free energy using the Feynman rules.  }

\newpage

\setcounter{page}{1}
  
\section{Introduction}
\label{sec:Introduction}

In this paper, which is a continuation of \cite{Kostov:1999xi}, we
formulate the $1/N$ expansion of $U(N)$ invariant matrix integrals as
the quasiclassical expansion in conformal field theories on Riemann
surfaces.  Our basic example will be the hermitian matrix model
defined by the partition function
\be\la{Coulgaz}
\CZ_N[V]  
&=& \int d\M\, e^{-\Tr V(\M)}
\no \\
&\sim & \int\limits _ \CC \prod _{i=1}^N d\l_i \ e^{- V(\l_i)} \
\prod_{i< j} (\l_i-\l_j) ^2.  \ee
Here $z_1,\dots, z_N$ are the eigenvalues of the matrix variable 
$\M$
and 
\be\la{defV}
 V(z) = - \sum_{n=0}^\infty t_n z^n 
 \, 
\ee
is a confining potential.  For the hermitian matrix integral the
contour of integration goes along the real axis.  Then the integral is
convergent if the potential becomes infinitely large when
$z\to\pm\infty$.  If this is not the case, the integral can be made
convergent by deforming the contour of integration.

The $1/N$ expansion of the free energy $\CF_N[V] = \ln\CZ_N[V]$ is of
the form
\be\la{Freetot} \CF_N[V] = \sum _{g=0}^{\infty} N^{2-2g}
\CF^{(g)}[V/N] + \ {\rm nonperturbative\ terms}, \ee
where the term $\CF^{(g)}$ counts the number of genus $g$ fat Feynman
graphs.\footnote{Strictly speaking, this is true only in the case of
one-cut solutions.} The leading, genus zero, term can be obtained by
evaluating the saddle point eigenvalue distribution
\cite{Brezin:1977sv}.  In this approximation the eigenvalues can be
considered as a continuous charged liquid defined by the spectral
density \be\la{defrho} \rho(z) =\lim_{N\to\infty} {1\over N} \sum
_{i=1}^N \delta(z-z_i).  \ee
  
The collective field theory for the spectral density \re{defrho} is
ill defined at small distances and cannot be used to compute the
higher terms of the quasiclassical expansion.  This can be done by
solving the Ward identities for the integral \re{Coulgaz}, known also
as loop equations.  An improved iterative scheme for calculating the
higher genus observables, known as method of moments, was set up by
Ambjorn, Chekhov, Kristjansen and Makeenko \cite{Ambjorn:1992gw}.
More recently, Eynard \cite{Eynard:2004mh} formulated the recursion
procedure for solving the loop equations in an elegant graphical
scheme.

The form of the loop equations suggests that there is a local field
theory hidden behind.  Indeed, the loop equations can be formulated as
Virasoro constraints for the bosonic field
 \be \la{defphi} \phi (z)= \sqrt{2}\sum_{i=1}^N \log\( z-z_i\)
 -{1\over \sqrt{2}} V(z) \, .  \ee
Therefore the theory in question must be a two-dimensional chiral CFT.
An important feature of this CFT is that the bosonic field develops a
large classical expectation value $\phi_c(z)$.  The classical current
$J_c=\p_z\phi_c$ satisfies a quadratic equation, the classical
Virasoro constraint, whose solution defines for a general potential a
hyperelliptic Riemann surface with branch points $a_1, \dots , a_{2p}$
placed on the real axis.  The set of branch cuts along the segments
$[a_{ 1}, a_{2}],\dots, [a_{ 2p-1}, a_{2p}]$ coincide with the support
of the classical spectral density.  Each branch cut is associated with
a local minimum of the potential and is characterized by the filling
number $N_j$, equal to the number of the eigenvalues trapped around
this minimum.

In the chiral CFT the branch points of the Riemann surface can be
thought of as the result of applying {\it twist} operators to the
$sl(2)$ invariant vacuum.  The twist operators are however not
conformal invariant.  In particular the translations change the
positions of the branch points.  In \cite{Kostov:1999xi} we claimed
that in order to satisfy the conformal Ward identity the twist
operators should be dressed by the modes of the twisted boson.  The
dressed twist operators were called in \cite{Kostov:1999xi} star
operators because of the analogy with the star operators introduced by
G. Moore in \cite{Moore:1990aa}.  We should stress here that the star
operators in \cite{Kostov:1999xi} and in \cite{Moore:1990aa} are
different objects.

The problem of computing the genus expansion of the of the observables
is thus reduced to the perturbative construction of the operators
dressing the twist fields.  Such  dressing
operator  represents a formal series expansion in the modes of
the twisted boson.  The coefficients of  the  expansion 
are related to the correlation functions in the
Kontsevich model.  In this sense the proposal made in
\cite{Kostov:1999xi} allows to decompose the hermitian matrix model
with an arbitrary potential into $2p$ Kontsevich models coupled
through the modes of a gaussian field on the hyperelliptic surface.
Similar decomposition formulas for the hermitian matrix model were
suggested in \cite{A.Alexandrov:2009aa,
Alexandrov:2008yq, A.Alexandrov:2007aa}.

The aim of this paper is to present the details of the construction of
the dressed twist operators and to set up the Feynman diagram
technique for calculating the free energy and the expectation values
in all orders on $1/N$.

The paper is organized as follows.  In Section \ref{section:2} we
remind the Fock space representation of the integral \re{Coulgaz} in
terms of a chiral bosonic field $\phi(z)$ with Liouville-like
interaction given in \cite{Kostov:1999xi} and derive the loop
equations from the conformal Ward identity.  This Section helps to
make the presentation self-consistent and can be skipped by the reader
who is familiar with the loop equations.  In \ref{section:3} we give
the representation of the leading and the subleading orders of the
free energy in terms of the correlation function of $2p$ twist
operators.  In Section \ref{section:4} we give the explicit
prescription for dressing the twist operators and set up the Feynman
rules for computing all orders of the genus expansion.  This Feynman
diagram technique is different than the diagram technique in
\cite{Eynard:2004mh}.  In Section \ref{section:5} we compute, using
the Feynman rules, the genus two free energy and compare with the
result of \cite{Ambjorn:1992gw}.

   \section{ The matrix model as a chiral CFT}
\label{section:2}

    \subsection{From Coulomb gas to compactified chiral boson}
 
The Coulomb gas integral \re{Coulgaz} can be represented as a Fock
space expectation value in a theory of a Liouville-like chiral CFT
with the wrong sign of the kinetic term.  Introduce the holomorphic
scalar field $\phi(z) $ with mode expansion at $z=\infty$
\be\la{boZmodes} \phi (z) &=& \hat q + J_0 \ln z - \sum _{n\ne 0} {J
_{n} \over n } z^{-n} \ ,\quad [J _n,J _m] = n \, \delta_{n+m, 0} ,
\quad [J_0 , \hat q ] = 1\, , \ee
and left and right Fock vacua defined by
\be\la{defFockv} \langle 0 | J _0 = \hat q |0\rangle = 0 , \quad
\langle 0 | J_{-n} = J _{n} |0\rangle =0 \quad (n\ge 1) .  \ee The
field has operator product expansion
\be\la{OPEvp} \phi (z) \, \phi (z') \sim \ln (z-z')\, \ee
and give a representation of the $\hat su(2)$ current algebra:

  \be\la{sutwoc} J = {\textstyle{1\over \sqrt{2}}} \ \p\phi(z) , \quad
  J_{\pm}= : e^{\pm \sqrt{2}\phi}:\, .  \ee
Then the partition function \re{Coulgaz} is given by the scalar
product
	\be\la{fockrp} \CZ_{N}= \langle N| e^{J [V]} \ e^{Q_+} |0
	\rangle, \ee
where
\be\la{nmltn} J[V] &=& -\oint\limits_\infty {dz\over 2\pi i} V (z)
J(z) = \sum_{n\ge 0} t_n \, J_n\, , \\
Q_+ &=& \int_{-\infty}^{\infty} d\l\ J_+(\l ) \ee
and $\langle N| $ is the charged left vacuum state
\be\la{defNvac} \langle N| = \langle 0| e^{\sqrt{2} N \hat q } \, .
\ee
The operator $e^{Q_+}$ generates screening charges, the operator
$e^{J[V]}$ produces the measure $e^{-V(z)}$ for each charge, and the
left vacuum $\langle N|$ projects onto the sector with exactly $N$
screening charges.  Similar Fock space representations of the
eigenvalue integral have been proposed in \cite{Marshakov:1991gc,
Kharchev:1992iv, Morozov:1995pb}.

The currents $J, J+$ and $J_-$ are invariant under the discrete
translations of of the field $ \phi$:
\be\la{perio} \phi \to \phi + i {\pi}{\sqrt{2}}.  \ee
Therefore the the field $\phi$ can be compactified at the self-dual
radius $R_{\rm s.d.}= 1/\sqrt{2}$.  Furthermore, the transformation
\be\la{orbif} \phi\to-\phi \ee
is an automorphism of the current algebra.  The geometrical meaning of
the symmetry \re{orbif} will become clear when we consider the
quasiclassical limit of the bosonic field.

 The correlation functions of in the matrix model are obtained through
 the identification
\be\la{resdeT} \phi (z)= \sqrt{2}\ \Tr\log\( z-\M\) -{1\over \sqrt{2}}
V(z) \, \ee
 or, in terms of the $\hat{su}(2)$ currents,
\be\la{resJ} J(z)= \Tr{1\over z-\M} - \hf V'(z) \, , \quad J_\pm (z)=
e^{ \mp V(z)} [ \det(z-\M)]^{\pm 2} .  \ee

\subsection{Conformal Ward identity}

 To prove that the theory is conformal invariant we have to
 demonstrate that the energy-momentum tensor
\be\la{vIIr} T(z) = \hf : \p\phi(z) \p\phi(z): \, =\sum_{n}L_n
z^{-n-2} \ee
 commutes with the screening operator $Q_+$.
   Indeed, for any $n\ge -1$ we have
 \be\la{vIr}
 [L_n, Q_+] = \int_{-\infty}^\infty d\l [L_n, 
J_+(\l)] =
  \int_{-\infty}^\infty d\l \, {d\over d\l} 
\Big(\l^{n+1} 
J_+(\l)\Big).
\ee
Since our potential diverges at infinity, the boundary terms vanish
and the result is zero for all $n\ge -1$.  As a consequence, the
expectation value
     \be \< T(z) \> \defeq \langle N| e^{J [V]} \ T(z) \, e^{Q_+} |0
     \rangle\, \ee
    is regular for $z\ne\infty$.  This condition can be written as a
    contour integral which projects to the positive part of the
    Laurent expansion of $T(x)$:
  \be \la{eqct} \oint\limits_{\infty} {dz'\over 2\pi i}\Big\langle{
  T(z) - T(z') \over z-z'} \Big\rangle =0.  \ee

The conformal Ward identity \re{eqct} is translated into a set of
differential Virasoro constraints on the partition function using the
representation of the gaussian field as a differential operator acting
on the partition function,
\be\la{PPhi} \phi(z)\ \to\ {\hat \phi} (z) \defeq{1\over \sqrt{2}} \,
\sum _{n \ge 0} t_n z^{n} + \sqrt{2}\, \ln z\, {\p\over \p t_0} +
\sqrt{2} \sum _{n\ge 0} {z^{-n}\over n} {\p\over \p t_n} .  \ee
    The Virasoro constraints read
  \be\la{viraa} {\hat L}_n\cdot \CZ_N=0 \ \ \ \ \ \ ( n\ge -1) \ee
  where \be\la{lmOm} {\hat L}_n\defeq\sum_{k=0}^{n} {\partial\over
  \partial t_k} {\partial\over \partial t_{n-k}}+ \sum_{k=0 }^{\infty}
  k \, t_k {\partial\over \partial t_{n+k}}, \qquad {\p \over \p t_0}
  \CZ_N = N \CZ_N. \ee

 \section{ The  quasiclassical limit}
 
 \label{section:3}

\subsection{The classical solution as a hyperelliptic curve}

Applied to the genus expansion \re{Freetot} of the free energy, the
Virasoro constraints \re{viraa} generate an infinite set of equations
for the correlation functions of the current $J(z)$, which can be
solved order by order in $1/N$.  The lowest equation is the classical
Virasoro condition, which determines the expectation value $J_c(z)$ of
the current in the large $N$ limit.  In our normalization $J_c$ is of
order of $N$, just as the confining potential $V$.

  The classical Virasoro constraint states that $T_c = J_c^2$ is an
  entire function of $z$.  The most general solution is
\be\la{clpp} J_c(z) =- M(z) \, y(z), \quad y^2= \prod_{j =1}^{2p}
(z-a_j) .  \ee
where $M(z)$ is an entire function of $z$.  Assuming that
$a_1<a_2<...<a_{2p}$, the meromorphic function $J_c(z)$ is
discontinuous along the intervals $[a_{2k-1},a_{2k}]$ and its
discontinuity is related to the normalized classical spectral density
by
    \be\la{clrho} \rho_c(x) = {J_c(x-i0)- J_c(x+i0) \over 2\pi i N} \,
    .  \ee

Our aim is to construct a CFT associated with the classical solution
$J_c$.  It is advantageous to think of the gaussian field as defined
not on the complex plane cut along the intervals $[a_{2k-1},a_{2k}]$,
but on the Riemann surface representing a two-fold branched cover of
the complex plane, the two sheets of which are sewed along the cuts
$[a_{2j-1}, a_{2j}]$.  In this way we trade the boundary condition
along the cuts for the monodromy relations ($\phi\to -\phi$) when one
moves around the branch points.

\begin{figure} 
\begin{center}
\includegraphics[width=200pt]{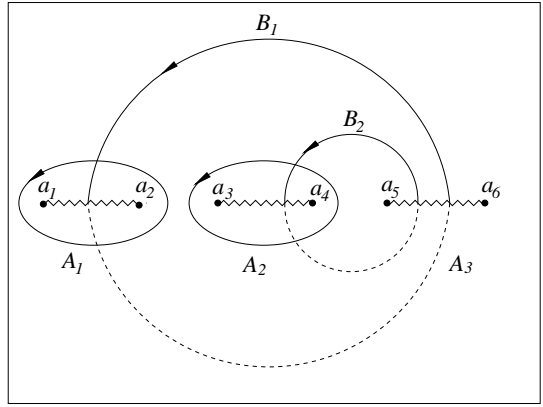} 
\end{center}
 \caption{ \small The $A$ and $B$ canonical cycles for a genus 2
 spectral curve $(p=3)$}
\label{cts}
\end{figure}

The hyperelliptic Riemann surface is characterized by a set of moduli,
associated with the canonical $A$ and $B$ cycles.  The cycle $A_j$
encircles the cut $[a_{2j-1},a_{2j}]$ and the cycle $B_j$ encircles
the points $a_{2j}, ...  ,a_{2p-3}$, passing through the $j$-th and
the $p$-th cuts (Fig.~\ref{cts}), so that
\be\la{eeqb}A_k \circ B_j = \delta _{k,j} \qquad (j=1, \dots, p-1).
\ee
  The classical current is determined completely by the potential
  $V(z)$ through the asymptotics
 \be\la{asymHc} J_c(z)= - \hf V'(z) + {N}{z^{-1}}+...  \ee
 and by the charges $N_1+\dots +N_{p}=N$ associated with the
 $A$-cycles,
  \be\la{normMC} \oint _{A_j} {dz\over 2\pi i}\ J_c(z) =N_j, \quad
  j=1,..., p.  \ee
	From \re{normMC} it follows that the derivatives
  \be\la{abdIf} \o_j = { \p J_c(z)\over\p {N_j} } dz \ee
     form a  basis of  holomorphic   differentials  of first kind,
  associated with the  cycles  $A_k$, 
\be\la{abdiF}
\frac{1}{2\pi i}\oint_{A_k}   \o_j  =   \delta_{kj} 
  .
  \ee
The integrals of $\o_j$ along the $B$-cycles give the period matrix of
the hyperelliptic curve
\be\la{eeqa}
 \tau_{kj} = {1\over 2\pi i} \oint_{B_k}  \o_j
.
\ee
    
In the following we will consider the filling numbers $N_j$ as fixed
external parameters.  The corresponding free energy is that of a
metastable state, which becomes stable up to exponentially small
effects in the large $N$ limit.  We will understand the $1/N$
expansion \re{Freetot} of the free energy in this setting.
Alternatively, one can introduce chemical potentials $\Gamma_j$ for
the charge $N_j$ and evaluate the partition function for fixed
$\Gamma_j$.

If one is interested in the quasiclassical evaluation of the original
partition function \re{Coulgaz}, one should perform the sum over all
possible $N_j$.  This problem has been considered and solved in
\cite{Bonnet:2000dz}.  After performing the sum, the logarithm of the
partition function does not have the $1/N$ expansion \re{Freetot}.
Since the numbers $N_j$ are the discontinuities of the bosonic field
along the $A$-cycles, the sum over all $N_j$ means that the bosonic
field is effectively compactified at the selfdual radius.

   \subsection{The branch points as primary conformal fields}

Our goal is to construct a CFT associated with the classical solution
$J_c$.  We will think of the hyperelliptic Riemann surface as the
complex plane with conformal operators, the {\it twist operators }
with dimension 1/16, associated with the branch points.  This point of
view was first advocated by Alexei Zamolodchikov in
\cite{Zamolodchikov:1987ae}.  Later some of the findings of
\cite{Zamolodchikov:1987ae} were obtained independently by Dixon {\it
at al} \cite{Dixon:1986qv} and further developed by a series of
brilliant papers by V. Knizhnik
\cite{Knizhnik:1986kf,Knizhnik:1987xp,Knizhnik:1989ak}.

Let $a$ be one of the branch points of the classical solution $J_c $.
In the vicinity of $a$ the current has mode expansion
\be\la{modexJ} J(z) = J_c(z) + \sum_{r\in \IZ +{1\over 2}} J_r \ (z-a)
^{-r -1} \ee
with the following algebra,
\be\la{comrr}
[J_r  , J_{s} ] = \hf  r \d_{r+ s, 0}.
\ee

The Ramond vacuum associated with this branch point is defined as the
highest weight vector of the representation of this algebra.  The
corresponding quantum field is the twist operator $\s(a)$
\be
  J_{1/2}   \s(a)     =J_{3/2}   \s(a)= J_{5/2}   \s(a)   = \dots 
  =    0   .
  \ee
The Hilbert space associated with $\s(a)$ generated by multiple action
on $\s(a)$ with the negative modes $J_{-1/2}, J_{-3/2}, \dots $ of the
current $J$.
   
The product of two twist operators $\s(a_1)\s(a_2)$ is a single-valued
operator with respect to the current $J(z)$.  This state can be
decomposed into charged eigenstates \re{normMC}.  Let us denote a
state with given charge $N$ by $[\s(a_1)\s(a_2)]_N$.  Our strategy in
the following is to simulate the cuts $[a_{2j-1}, a_{2j}]$ of the
Riemann surface by multiplying the right vacuum with the operators
$[\s(a_{2j-1}) \s(a_{2j})]_{N_j}$ and give a Fock-space representation
of the partition function of the matrix model similar to \re{fockrp}.
In the new representation the operator $e^{Q_+}$ creating the $N$
charges is replaced by a product of $p $ pairs of twist operators.

The correlation function of $p $ pairs of twist operators was
calculated by Al.  Zamolodchikov \cite{Zamolodchikov:1987ae} in the
case when the current has no expectation value and the total charge is
zero:
\be \big< 0\big | \, \prod_{j=1}^{p } [\s(a_{2j-1}) \s(a_{2j})]_{N_j}
\big|0\big> =Z_{\text{twist}}(a_1,\dots, a_{2p}) \ e^{ i\pi \sum _{j,
k=1}^p \t_{jk} N_j N_k}\ .  \ee
The meromorphic function $Z_{\text{twist}}(a_1,\dots, a_{2p})$ is
given by
   \be\la{defZl} Z_{\text{twist}}= \prod_{j<k} ^{2p} \( a_j-a_k\)^{-
   1/8} [\det K]^{-1/2} , \ee
where the matrix $K$ is defined by
   \be\la{defK}
 K_{ij} =  \int_{A_i} {z^{j-1} dz\over y(z)}
 \qquad (i,j = 1,\dots, p-1).
 \ee
It is straightforward to generalize this formula to the case of
non-zero total charge and non-vanishing expectation value of the
current.  We should simply insert the operator $\exp J[V]$ as in
\re{fockrp}.  This gives a representation of the partition function in
the quasiclassical limit as the scalar product
\be \la{firsttwo} \CZ_N^{\rm quasicl } = \ \big< N \big | \, e^{ J[V]}
\, \prod_{j=1}^{p } [\s(a_{2j-1}) \s(a_{2j})]_{N_j} \big|0\big>.  \ee

The normalized expectation value of the current evaluated with respect
to \re{firsttwo} obviously coincides with the classical solution
$J_c$:
 \be \< J(z)\>_{\rm quasicl } \defeq \sqrti\  \p \hat \phi(z) \cdot \log
 \CZ_N^{\rm quasicl }= J_c(z)\, .  \ee
Therefore the r.h.s. of \re{firsttwo} satisfies the classical Virasoro
constraints and  reproduces correctly the leading order of
the free energy.  The quantum Virasoro constraints are however not
respected.  The twist operator $\s(a)$ depends on the position of the
branch point $a$ and does not satisfy the lowest Virasoro condition
$L_{-1}$.  It is also not invariant with respect to dilatations
generated by $L_{0}$.

 \section{The genus  expansion}
\label{section:4}

 \subsection{Dressed  twist operators}

We would like to modify \re{firsttwo} so that it reproduces all orders
of the $1/N$ expansion of the free energy.  We have to look for
operators which create conformally invariant states near the branch
points.  Such operators can be constructed from the modes of the
twisted bosonic field near the branch point by requiring that the
singular terms with their OPE with the energy-momentum tensor vanish.

In \cite{Kostov:1999xi} the author proposed\footnote{This proposal was
reproduced in more details in sect.  4.3 of \cite{Dijkgraaf:2004ac}
following an unpublished extended version of \cite{Kostov:1999xi}.  }
that twist operator $\s(a)$ can be made conformal invariant by
multiplying with an appropriate dressing operator $e^{\hat w(a)}$,
which is made from the modes of the twisted field:
  \be\la{Starop} \sigma(a) \to S(a) = e^{\hat w(a)}\sigma(a) .  \ee

Assuming that the dressed twist operator is a well defined operator in
the Hilbert space associated with $ \s(a)$, the dressing exponent
$w(a)$ can be expanded as a formal series in the creation operators
$J_{-r} \ (r>0)$ defined by the expansion \re{modexJ} near the point
$a$:
\be\la{expFf} \hat w(a) = \sum _{n\ge 0} {1\over n!}
\sum_{r_1,...,r_n} w_{ r_1...r_n} (a)\, J_{-r_1}...  J_{-r_n}.  \ee
The coefficients of the expansion depend on the classical current
$J_c$ and are determined by the requirement of conformal invariance
  \be\la{vViI} L_n \ S(a) = 0 \qquad (n\ge -1) \ee
where the Virasoro generators $L_n$ are defined by the expansion at
the point $z=a$ of the energy-momentum tensor
   \be\la{defTz} T(z) = \lim\limits_{z'\to 0} \[ J(z)J(z') - {1\over 2
   (z-z')^2}\]= \sum_{n} L_n (a) \ (z-a)^{-n-2}.  \ee
 One finds for the Virasoro operators with $n\ge -1$
\be\la{lmOm1}
 L_n
&=&   \sum_{r+s=n} :\left(J_r + J^c_{r}   \right)
\left( J_{s}+J^c_{s} 
\right):
+ {1\over 16  }\delta_{n,0}  
 ,
\ee
 where $J^c_r$ are the modes in the expansion of the classical current
 near the point $z=a$:
   \be\la{defmus}
   J_c(z) = \sum _{r\ge 3/2}  J^c_{-r}  \, (z-a)^{r-1}.
   \ee
   By convention  $J^c_{-r} =0$ if $r\le 1/2$.
   
Once we found conformal invariant operators that create the branch
points, it is clear how to repair \re{firsttwo} so that it holds for
all orders in $1/N^2$.  It is still possible to assign to the pair of
dressed twist operators associated with the endpoints of the cut
$[a_{2j-1} ,a_{2j}]$ a definite charge $N_j$.  We denote this state by
$ [S(a_{2j-1}) S(a_{2j})]_{N_j}$.

 Our claim is that the partition function of the matrix model is
 equal, up to non-perturbative terms, to the expectation value
 \be \la{operS} 
 \CZ_N
=
\ \big< N  \big |   \,  e^{ J[V]}  \, \prod_{j=1}^{p }
  [S(a_{2j-1}) S(a_{2j})]_{N_j}
\big|0\big>,
\ee
 where $J(z)$ is the current of the $\IZ_2$-twisted gaussian field
 defined on the Riemann surface of the classical solution.  Indeed,
 this expression satisfies the conformal Ward identity in all orders
 of $1/N$ since by construction the energy-momentum tensor \re{defTz}
 commutes with the dressed twist operators.  Furthermore, the leading
 order expectation value of the current $ J_c(z) $ satisfies the
 asymptotics at infinity \re{asymHc} and the conditions \re{normMC}.

 \bigskip
 
 {\it Remark:} \ The Fock space realization \re{operS} of the
 partition function resembles the QFT representation of the
 $\tau$-function for isomonodromic deformations obtained by T. Miwa
 \cite{Miwa:1980tu} and revisited by G. Moore \cite{Moore:1990aa}.  In
 our case there is an irregular singularity at infinity and $2p$
 regular singularities at the branch points.  For matrix ensembles
 with hard edges (infinite wall potential), it was convincingly argued
 in \cite{Gaberdiel:2005aa} that the Fock space representation
 \re{fockrp} can be transformed into to a correlation function of
 Moore's star operators.  However, even in this simplest case, the
 perturbative, or $1/N$, expansion of the star operators is ill
 defined.  The ambiguity gets even worse for smooth potentials.  The
 star operator is defined in \cite{Moore:1990aa} by the exponential of
 a contour integral starting at the branch point, but in the case of a
 smooth potential the position of the branch point should be adjusted
 at each order in $1/N$.  The dressed twist operators \re{Starop}
 generate the complete perturbative expansion, while the star
 operators of \cite{Moore:1990aa} capture the leading non-perturbative
 behavior.

\subsection{Fock space representation of the partition function}

The $1/N$ expansion can be obtained by considering the dressing
operators as a perturbation and expand them in the negative modes of
the current.  Let us denote by $J_r^{[a_j] }$ the modes of the current
$J$ associated with the expansion around the branch point $a_j$ .
Then the total dressing operator, which we denote by $ \hat\Omega$, is
given by the formal series
 \be\la{exphatO} \hat\Omega =\prod_{ j=1, \dots, 2p} e^{\hat w(a_j)},
 \quad \hat w(a_j)= \sum _{n\ge 0} {1\over n!} \sum_{r_1,...,r_n}
 w^{[a_j]} _{ r_1...r_n} \ J^{[a_j]}_{-r_1}...  J^{[a_j]}_{-r_n}\, .
 \ee
The partition function $\CZ_N$ is equal, up to non-perturbative
corrections, to the normalized expectation value of the dressing
operator with respect to the left and right states
\be \big< \text{left} \big| \defeq \big< N \big | \, e^{ J[V]} ,
\qquad \big| \text{right} \big> \defeq \prod_{j=1}^{p } [\s(a_{2j-1})
\s(a_{2j})]_{N_j} \big|0\big>.  \ee

In order to perform the expansion we should evaluate the expectation
value of any product of negative modes $J_r^{[a_j] }$.  Since $J(z)$
is the current of a gaussian field, it is sufficient to calculate the
expectation value of a pair of such modes:
  \be\la{defpropG} G_{r, s}^{[a_i, a_j]} = \langle J_{-r}^{[a_i]}\
  J_{-s}^{[a_j]} \rangle \, \defeq\, { \big< \text{left} \big| \,
  J_{-r}^{[a_i] }\ J_{-s}^{[a_j] }\, \big| \text{right} \big> \over
  \big< \text{left} \big| \text{right} \big> }.  \ee
 The matrix $G^{(a_i , a_j)}_{r, r'} $ can be computed knowing the
 two-point function $\< J(z)J(z')\>$, which is the unique function
 defined globally on the Riemann surface and having a double pole at
 $z=z'$ with residue $1/2$.  From the definition \re{defpropG} and the
 mode expansion \re{modexJ} it follows that
\be\la{gijaa} G^{[a_i a_j]}_{r, r' }= \int {dz\over 2\pi i } \int
{dz'\over 2\pi i} \ { \langle J(z) J(z') \rangle \over (z-a_i )^r (
z'-a_j )^{r'}} \, .  \ee
Once we know the matrix $ G_{r, s}^{(a_i, a_j)} $ and the coefficients
$ w^{[a_j] } _{ r_1...r_n} $, we can compute the $1/N$ expansion to
any order just by expanding the dressing operators and performing Wick
contractions.

This prescription can be packed in a concise formula in the following
way.  Introduce, together with the right Fock vacuum associated with
the $2p$ twist operators, a left twisted Fock vacuum, which
annihilates the negative modes of the current.  The two vacuum states,
which we denote by $\ltwv $ and $\rtwv $, are defined by
 \be \ltwv \, J_{-r}^{[a_j] } = 0, \ \ J_{r}^{[a_j] } \rtwv = 0 \qquad
 (r\ge 1/2; \, j=1, \dots, 2p).  \ee
Then the state $ \big| \text{right} \big> $ can be identified with
$\rtwv$ and the state \big< \text{left} \big| is obtained from $\ltwv$
by acting with the gaussian operator associated with the matrix
\re{defpropG}.  As a result we obtain the following Fock space
representation of the expectation value \re{operS},
\be\la{FockZ} \CZ_N= \big< \text{left} \big| \hat\O \big| \text{right}
\big> = \big< \text{left} \big| \text{right} \big>\ \ltwv \ e^{2 J
\hat G J} \ \hat\Omega \ \rtwv\, , \ee
where $\hat \O $ is defined by \re{exphatO}
and 
\be\la{defJGJ} J \hat G J \ \defeq\ \sum_{i,j =1}^{2p} \sum_{r, s \ge
1/2} {1\over r\, s} G_{r, s}^{[a_i, a_j]} J_{r}^{[a_i] }\,
J_{s}^{[a_j] }.  \ee

\subsection{The dressing operator and the  Kontsevich integral}

In order to make use of this formula we need the explicit expressions
for the coefficients of the series \re{expFf}, which can be obtained
by demanding that the dressing operator $\hat\O = e^{\hat w}$ solve
the Virasoro constraints generated by the operators \re{lmOm1}.

A detailed analysis of the solution of these Virasoro constraints can
be found in \cite{Itzykson:1992aa}.  The solution is given by the
Kontsevich matrix integral \cite{Kontsevich:1992ti}, also known as
matrix Airy function.  To make the connection with the notations of
\cite{Itzykson:1992aa} we represent the modes $J_r$ as\footnote{The
times $t_n$ here are shifted with respect to the times in
\cite{Itzykson:1992aa} as $(t_n)_{\text{here}}=[(2n-1)!!]^{-1}( t_n
-\d_{n, 1})_{\text{there}}.  $ }
\be\la{ReprJt} J_{-n- 1/2} = - \hf { t_n } \, , \quad J_{n+1/2} = - {
(n+\hf) } \ \p_n \hskip 1cm (n\ge 0) \ee
where we denoted $\p_n \equiv \p/\p t_n$.  We also relabel the modes
of the classical current as\footnote{The moments $I_n$ of
\cite{Itzykson:1992aa} are related to $\mu_n$ by $ I_1 =1- \mu_1$ and
$I_n = - (2n-1)!!  \, \mu_n$ for $n\ge 2$.}
\be
J^c_{-n-{1\over 2}} \ \to - \hf \mm_n  \hskip 1cm  (n\ge 0).
\ee
Then the dressing operator \re{expFf} is represented by a function
$\O(t_0, t_1,t_2,\dots)$ which satisfies
\be
\hat L_n \, \O = 0 \quad (n\ge -1)
\ee
 with  
\be\la{Virtn} \hat L_n&=& \sum_{k-m= n}(k+\hf)( t_{m} + \mm_m)\p_k +
\sum_{k+m=n-1} (k+\hf)(m+\hf) \p_k\p_m\\
&+&      
 {1\over 4}\,  t_0^2\,  \d_{n+1,0}+{1\over 16} \d_{n,0}
 \hskip 1cm  (n\ge -1).
 \ee

 The solution depends on the moments $\mu_n$ of the classical current.
 It is sufficient to have the solution $\O_0$ for the simplest
 nontrivial classical background $\mm _n = \mm _1 \, \d_{n,1} $.  Then
 the general solution is obtained simply by a shift $t_n\to t_n
 -\mm_n$, $n\ge 2$:
\be \O =\exp\Big(- \sum_{n\ge 2} \mm_n \p_n\Big)\ \O_0 .  \ee

The function $\O_0$ is given by a formal expansion in $t_0, t_1, \dots
$ and $1/\mm _1\sim 1/N$:
  \be\la{expOmo} \O_0(t_n) = {\mm _1} ^{-1/24} \exp \sum_ {g\ge 0} \
  \sum_{n\ge0}\ \sum_{k_1,\dots, k_n\ge 0}{ \mm _1} ^{2-2g-n} \
  w^{(g)}_{k_1, \dots, k_n}\ {t_{k_1}\dots t_{k_n}\over n!}.  \ee
The coefficients $ w^{(g)}_{k_1, \dots, k_n}$ are the genus $g$
correlation functions in the Kontsevich model.  They are proportional
to the intersection numbers in the moduli space of Riemann surfaces of
genus $g$ with $n$ punctures:
 \be\la{wtaucon} w^{(g)}_{k_1, \dots, k_n} = (-1)^n \prod_{j=1}^n
 (2k_j-1)!!\ \langle \t_{k_1} \dots \t_{k_n}\rangle_g .  \ee

The intersection numbers $ \langle \t_{k_1} \dots \t_{k_n}\rangle_g$
are positive rationals.  They are nonzero only if the indices
$k_1,\dots , k_n$ obey the selection rule
  \be\la{gnus} \sum_{j=1}^n k_j = 3(g -1)+n .  \ee
The genus zero intersection numbers are the multinomial coefficients
 \be\la{ssts} \langle \tau_{m_1}...\tau_{m_n}\rangle _{_0} =
 {(m_1+...+m_n)!  \over m_1!...m_n!} , \ \ m_1+...+m_n= n-3.  \ee
The first several intersection numbers of genus $g =1 , 2 $ are
\cite{Itzykson:1992aa}
 \be\la{tauSS} &&\langle \tau_ 1^n \rangle_{_1 }={(n-1)!\over 24}, \
 \langle \tau_ 0^n\tau_{n+1} \rangle_{_1 }={1\over 24}, \ \langle\tau_
 0 \tau_ 1 \tau_ 2 \rangle_{_1 }={1\over 12}, \no\\
  &&\langle \tau_ 2 ^3 \rangle_{_2 }
= {7\over 240}, \  \langle \tau_ 2  \tau_3 \rangle_{_2 }= {29\over 
5760},\  \langle \tau_ 4\rangle_{_2 }
= {1\over 1152}.
\ee 
An efficient procedure for evaluating $ \langle \t_{k_1} \dots
\t_{k_n}\rangle_g$ was proposed in \cite{Bergere:2009aa}.

  \subsection{ Feynman rules}
  
Now we have all the ingredients needed to construct the $1/N$
expansion of the free energy associated with the classical solution
with $p$ cuts.  Our starting point is the Fock space representation
\re{FockZ} of the all genus partition function.  The expression for
the partition function depends on the moments of the classical current
at the branch points
 \be \mm _n^{[a_j] } = -2 \oint_{a_j} {d z\over 2\pi i} {J_c(z)\over
 (z-a_j)^{n+1/2}} \, \qquad (n\ge 1) \ee
and the matrix $g^{[a_i a_j]}_{m, n } $ associated with the two-point
function of the current on the Riemann surface and defined by
\be\la{gijaan} G^{[a_i a_j]}_{m, n }= 4 \int {dz\over 2\pi i } \int
{dz'\over 2\pi i} \ { \langle J(z) J(z') \rangle \over (z-a_i
)^{m+1/2} ( z'-a_j )^{n+1/2}} \, .  \ee

 With each branch point we associate a set of coordinates $t_n^{[a_j]
 }$ and use the representation \re{ReprJt} of the modes of the bosonic
 current in terms of the Heisenberg algebra generated by $t_n^{[a_j]
 }$ and $\p_n^{[a_j] }$.  Together with the explicit solution
 \re{expOmo} for the dressing operators associated with the branch
 points, this leads to the following expression for the genus
 expansion of the free energy:

 \be\la{treprF} && e^{N^{2 } \CF^{(0)} +\CF^{(1)} }= \ e^{ i\pi \sum
 _{j, k=1}^p \t_{jk} N_j N_k + \sum_n t_n J_n}\ \prod_{j=1}^{2p}
 \(\mu_1^{[a_j]}\)^{-1/24} Z_{\text{twist}}(a_1,\dots, a_{2p}) \, ;
 \quad \\
\no \\
&&e^{\sum_{g\ge 2}N^{2-2g} \CF^{(g)} } = \exp\( \hf \sum_{i,j=1}^{2p}
\sum _{m,n\ge 0 } G^{[a_i a_j]}_{m, n } \p_m^{[a_i] }\p_n^{[a_j] }\)
\exp\( \sum_{j=1}^{2p} \sum _{n\ge 0} \mm _n^{[a_j] }\p_n^{[a_j] }\)
\no\\
&&\qquad\qquad \times 
\exp\(   \sum_ {g\ge 0} \      \sum_{n\ge0}\ 
  \sum_{k_1,\dots, k_n\ge 0} \(\mm _1^{[a_j] }\)^{2-2g-n} \
  w^{(g)}_{k_1, \dots, k_n}\  {t_{k_1}^{[a_j] }\dots t_{k_n}^{[a_j] }\over n!}
\)_{t_{\cdot}^{(\cdot )} =0}. 
\quad \ee  

Using this formula one can evaluate the genus $g$ free energy as a sum
of a finite number of Feynman graphs with vertices
$w^{(g)}_{k_1,\dots, k_n}$ given by \re{wtaucon}, tadpoles $\mm
_n^{[a_j]}$ and propagators $G^{[a_i a_j]}_{m, n } $.  Note that while
the propagators and the tadpoles depend on the classical solution, the
vertices are universal.  From \re{tauSS} we find the first several
vertices \re{wtaucon}:
{\small \be && w^{(0)}_{0,0,0}= -1, \ \ w^{(1)}_{1}= -{1\over 24}, \ \
w^{(1)}_{1,1}= {1\over 24}, \ \ w^{(1)}_{0, 2}= - {1\over 8} , \ \
w^{(1)}_{0, 1,2}= -{1\over 4} , \ \ w^{(1)}_{0, 0,3}= - {5 \over 8} ,
\no\\
 && w^{(1)}_{0, 0,2,2}= {3\over 2}\ \ w^{(2)}_{2,2,2}= -{63\over 80},
 \ \ w^{(2)}_{2,3}= {29\over 128}, \ \ w^{(2)}_{4}= -{35\over 384} , \
 \ w^{(2)}_{0,0,0,0, 2}= -{3} .  \ee }

\begin{figure} 
\begin{center}
\includegraphics[width=375pt]{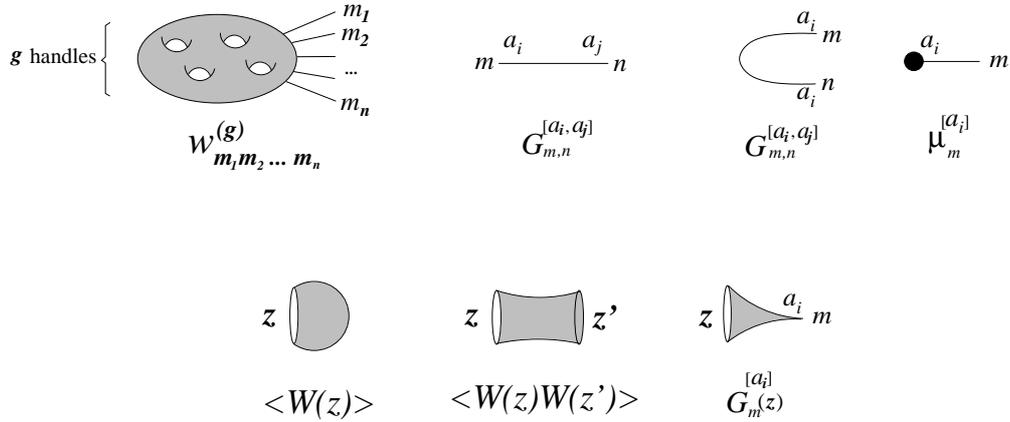} 
\end{center}
  \caption{\small{The Feynman rules for genus expansion of the
  hermitian matrix model.  The vertices $W^{(g)}_{m_1,\dots, m_n}$ are
  universal, the propagators $G^{[a_i,a_j]}_{m,n}$ depend only on the
  moduli of the spectral curve, and the tadpoles $\mm _m^{[a_j]}$ are
  determined by the expansion of the classical solution at the branch
  points.  The vertices are connected to each other by propagators and
  the tadpoles are connected directly to the vertices.  The Feynman
  graphs for the correlation functions of the resolvent contain three
  more elements: the disk, the cylinder and the external legs depicted
  in the second line.  }}
\label{Feynm}
\end{figure}

The correlators of the resolvent of the random matrix are obtained
from the correlations of the current $J$ by the identification
\re{resJ}.  To evaluate the correlation functions of the current we
use its representation as differential operator
  \be \hat J(z) = J_c(z) + \sum_{j=1}^{2p} \sum_{n\ge 0} G^{[a_j]
  }_n(z) \ \p^{[a_j] }_n.  \ee
 We therefore extend the set of Feynman rules by adding the external
 lines which represent the functions
 \be
 G^{[a_j] }_n(z)
 =  
2  \int {dz'\over 
2\pi i}  
\  { \langle    J(z) J(z') \rangle\over 
 ( z'-a_j )^{n+1/2}}
\, 
.
  \ee

The Feynman rues for evaluating the $1/N$ expansion are given in
Fig.~\ref{Feynm}.  The genus $g$ contribution for any observable is
equal to the sum of all genus $g$ Feynman diagrams.  The genus of a
Feynman diagram is equal to $2-2h-n$, where $h$ is the number of
handles, including the handles made of propagators, and $n$ is the
number of the external lines.

 \section{Example: the single cut solution }
  
 \la{section:5}

In this section we consider in details the case of one cut $(p=1)$,
where the Riemann surface of the classical solution is a sphere.  The
explicit expression for the classical current is
\be J_c(z) = \hf \oint_{A_1}{dz\over 2\pi i} {y(z)\over y(z')} \
{V'(z')-V'(z) \over z-z'}, \qquad y(z)= \sqrt{(z-a_1)(z-a_2)}.  \ee
Expanding at $z=\infty$ and using the asymptotics $J_c(z)\sim -{1\over
2} V'(z) + { N/z}+ ...$, one finds the conditions
  \be\la{strngeq} \oint_{A_1}{dz\over 2\pi i} { z^k V'(z) \over
  y(z)}=-2N \delta_{k,1}\quad\quad(k=0,1) \, \ee
which  determine the positions of the two branch points.

The two-point function of the $\IZ_2$-twisted current $J(z)$ on the
Riemann surface of the classical solution is equal to the 4-point
function of two currents and two twist operators:
 \be\la{prOp}
 \langle J(z)J(z')\rangle\equiv   \langle 
0|J(z)J(z')\sigma(a) \sigma(b)|0\rangle
 = { {\sqrt{(z-a_1)(z'-a_2)\over (z-a_2)(z'-a_1)}}+
 {\sqrt{(z'-a_1)(z-a_2) \over (z'-a_2)(z-a_1)}} \over 4(z-z')^2 } .
 \ee
  The coefficients $G ^{[a_i,a_k]}_{km} $ are obtained by expanding
 $\langle J(z)J(z')\rangle$ near the points $a_1$ and $a_2$.
 Assuming that $a_2>a_1$, we write 
 \be
 G ^{[a_i,a_j]}_{km}
 =
4 \oint _{a_i} {d z\over 2\pi i} 
 \oint _{a_j} {dz_j\over 2\pi i} \
  (z-a_i)^{-k -{1\over 2} } (z'-a_j)^{-m -{1\over 2} }
\  \langle J(z)J(z')\rangle .
 \ee
%
%
  We find that $ G ^{[a_1,a_1]}_{km}$ are of the form
\be\la{Gabb} G^{[a_i ,a_k]}_{k, m}= {1\over d^{k+m+1}}\ g^{[a_i ,
a_k]}_{km}, \qquad d= |a_1-a_2|, \ee
 where $g^{[a_i , a_j]}_{km}$ are rational numbers with the symmetry
 $g^{[a_1 , a_1]}_{km} = g^{[a_2 , a_2]}_{km} $ and $ g^{[a_1 ,
 a_2]}_{km} = g^{[a_2 , a_1]}_{km} $.  The first several coefficients
 $g^{[a_i, a_j]}_{km}$ are
\be\la{Gmns} \{ g^{(a_1,a_1)}_{km}\} _{_{k,m\ge 0} }= \{
g^{(a_2,a_2)}_{km}\} _{_{k,m\ge 0} } &=& {\begin{pmatrix} -{1/ 2} &
{3/ 8} & -{5/ 16} & ...\\
{3/ 8} & -{3/ 8} & {45/ 128} & ...\\
-{5/ 16} & {45/ 128} & -{45/ 128} &
  ...  \\
.\ .\ .  &.\ .\ .  &.\ .\ .  & \\
\end{pmatrix}
} \no\\
\{ g^{(a_1,a_2)}_{km}\} _{_{k,m\ge 0} }= \{ g^{(a_2,a_1)}_{km}\}
_{_{k,m\ge 0} } &=& {\begin{pmatrix} -{1 } & {3/ 2} & -{15/ 8} & ...\\
{3/2} & - 21/4 & 165/16 & ...\\
-15/8 & 165/16 & -1745/64& ...  \\
.\ .\ .  &.\ .\ .  &.\ .\ .  & \end{pmatrix}}.  \ee

 \begin{figure}
\begin{center}
\includegraphics[width=395pt]{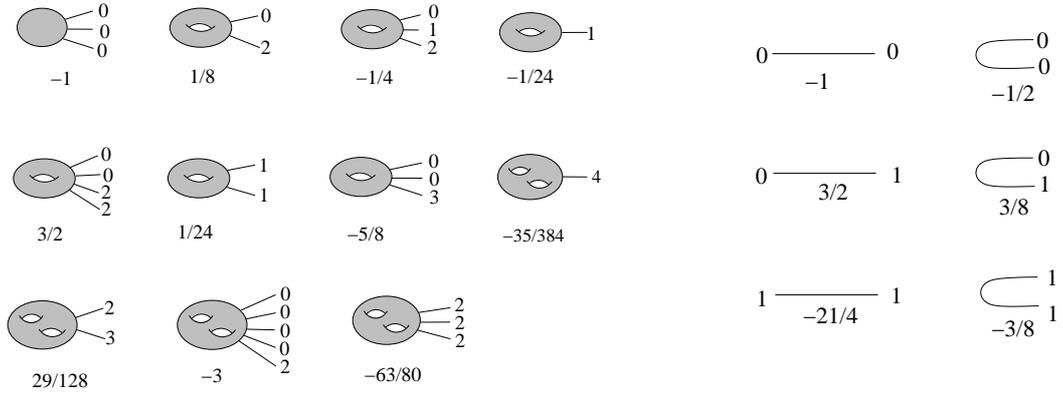}
\end{center}
 \caption{\small The vertices $w^{(g)}_{m_1,\dots, m_n}$ and the
 propagators $g^{[a, a']} _{mn}$ and $g^{[a, a]} _{mn}= g^{[a', a']}
 _{mn}$ contributing to the genus two free energy in the case of a
 single cut.}
\label{FeynmF2}
\end{figure}
\begin{figure}
\begin{center}
\includegraphics[width=390pt]{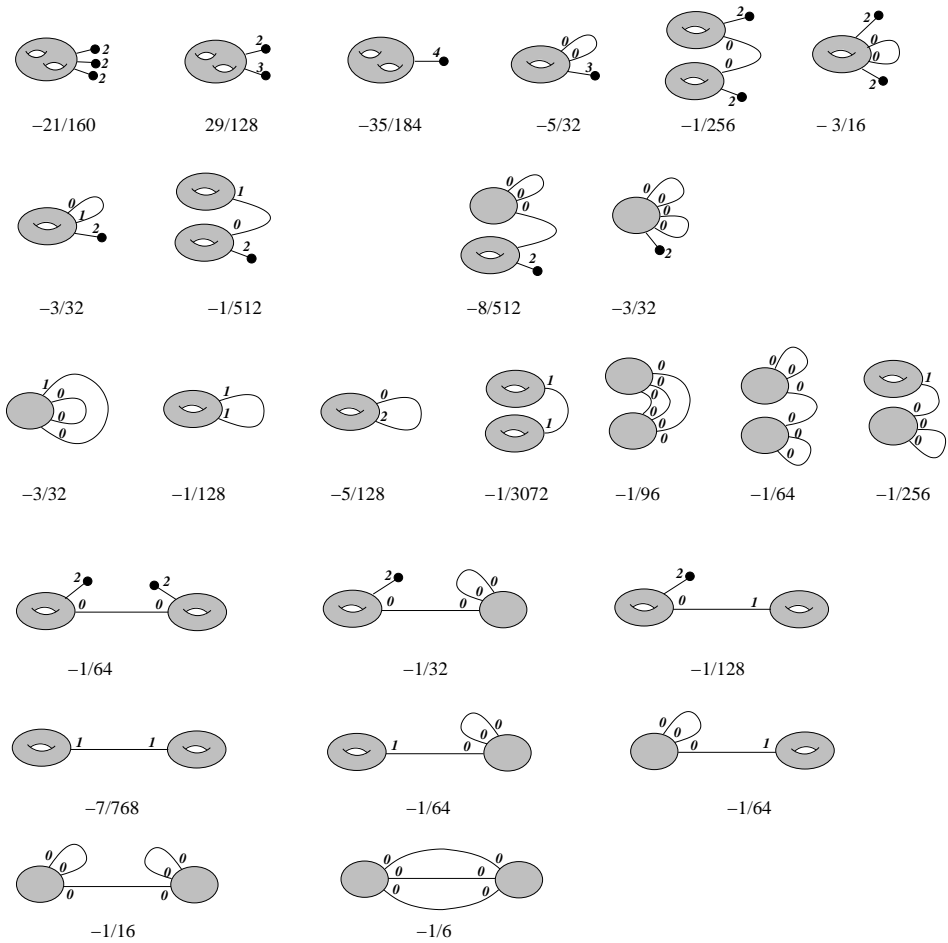}
\end{center}
 \caption{\small The Feynman graphs contributing to the genus two free
 energy in the case of a single cut.}
\label{DiagramsF2}
\end{figure}

As an illustration of the Feynman diagram technique we will evaluate
the free energy up to genus two.  We denote the two branch points by
$a'= a_1$, $a =a_2$, and the moments of the classical solution
associated with them by $ \mm _n = \mm _n^{(a)},\mm '_n = \mm
_n^{(a')}$.

The genus-one term is
\be \la{gnsone} \CF^{(1)} &=& -{1\over 24}\ln \mm _{1} -{1\over 24}\ln
\mm '_{1} - {1\over 8} \ln d\qquad (d=a-a') .  \ee
The first two terms come from the dressing of the twist operators and
the last term is the logarithm of the correlation function $\langle 0|
\s(a') \s(a)|0\rangle$.

The term $\CF^{(2)}$ in the genus expansion of the free energy is a
sum of the contributions of all possible genus two Feynman diagrams
composed by the vertices and the propagators shown in
Fig.~\ref{FeynmF2}.  The relevant diagrams are depicted in
Fig.~\ref{DiagramsF2}.  The result is\footnote{ The author thanks A.
Alexandrov for pointing out a missing term in the unpublished extended
version of \cite{Kostov:1999xi}.} {\small
\be\la{frEnr} \CF^{(2)} &=& {1\over \mm _1^2} \(- {21\over 160 }{\mm
_2^3\over \mm _1^3} +{29\over 128 } {\mm _2\mm _3 \over \mm _1^2} -
{35\over 384} {\mm _4\over \mm _1} + {5\over 32} {\mm _3\over \mm _1
d} - {49\over 256}{\mm _2^2\over \mm _1^2d} - {105\over 512} { \mm _2
\over d^2\mm _1} - {175 \over 1024}{1\over d^3} \) \no\\
& +& \{ \mm \leftrightarrow \mm '\}\ \ + {1\over \mm _1 \mm '_1 }
\left( -{1\over 64} {\mm _2\mm '_2\over \mm _1\mm '_1 d } - {5\over
128} {\mm _2\over \mm _1d^2} - {5\over 128} {\mm '_2\over \mm '_1d^2}
-{69\over 256} {1\over d^3} \right) \ee }

The genus two free energy was first computed by Ambjorn {\it et al} in
\cite{Ambjorn:1992gw}.  To compare with the result of
\cite{Ambjorn:1992gw} we have to express the moments $\mm _n$ and $\mm
'_n$ of the classical field in terms of the ACKM moments $M_n$ and
$J_n$, defined as {\small
\be\la{momentS} M_n &=& - {2\over N}\oint {dz\over 2\pi i} \ { J_c(z)
\over (z-a)^{n+1/2}(z-a')^{1/2}}\ = {1\over N } \sum_{k=0}^{n-1} {
(2k-1)!!  \over k!  (-2d)^k} \mm _{n-k} , , \no\\
 J_n &=& - {2\over N}\oint {dz\over 2\pi i} \ { J_c(z) \over
 (z-a)^{1/2}(z-a')^{n+1/2}} = ( -1)^{n-1} {1\over N} \sum_{k=0}^{n-1}
 { (2k-1)!!  \over k!  (-2d)^k} \mm '_{n-k}.  \ee }
Substituting \re{momentS} in the answer found in \cite{Ambjorn:1992gw}
(note that some of the coefficients are corrected in
\cite{ACKM-erratum})
%
{\small \be\la{ACKMa} N^2 \CF^{(2)} &= & - {181\over 480 J_1^2 d^4} -
{181\over 480 M_1^2 d^4} + {181 J_2\over 480 J_1^3 d^3} - {181
M_2\over 480 M_1^3 d^3} +{3J_2 \over 64 J_1^2 M_1 d^3}\\
& -&{3M_2 \over 64 M_1^2J_1 d^3} - {11 J_2^2\over 40 J_1^4 d^2} - {11
M_2^2\over 40 M_1^4 d^2}+ {43 M_3 \over 192 M_1^3 d^2}\\
&+& {43 J_3 \over 192 J_1^3 d^2} +{J_2M_2\over 64 J_1^2M_1^2d^2} - {5
\over 16 J_1M_1 d^4} + {21 J_2^3 \over 160 J_1^5 d} -{29 J_2J_3\over
128 J_1^4d} + {35 J_4\over 384 J_1^3 d}\\
& -&{21 M_2^3 \over 160 M_1^5 d} +{29 M_2M_3\over 128 M_1^4d} -{35
M_4\over 384 M_1^3d} \ee
 }
 one  reproduces   the expression  \re{frEnr}.

 \section{ Discussion}

  The CFT formalism was  developed before for the continuum
limit of a class of matrix models which reduce to  Coulomb gas  integrals  similar to \re{Coulgaz},
such as the SOS and ADE matrix models  \cite{Kostov:2007xw}. 
The classical solutions considered in  \cite{Kostov:2007xw}  correspond to 
non-hyperelliptic Riemann surfaces with one higher (or even infinite)  order branch point at 
infinity and one simple branch point on the first sheet.  The CFT formulation of these
models provides a rigorous derivation of the Feynman rules for the genus expansion 
 obtained previously  in \cite{Higuchi:1995pv,Kostov:1995xw}.

In this paper we developed the CFT formalism for any classical solution of the 
 hermitian one-matrix model. We do not see  conceptual difficulties to generalize 
 it to any classical solution of the above mentioned models, including the $O(n)$ model.
  
The Feynman rules that follow from the CFT representation look similar to
the diagram technique developed  in a larger class of matrix models by 
Eynard and collaborators 
\cite{Eynard:2004mh,Chekhov:2006vd,Chekhov:2006rq,Eynard:2008ac,Eynard:2008we}. 
If  there exists an exact correspondence between the two formalisms, 
then the CFT formalism can be possibly extended also for matrix models  which do not reduce to 
Coulomb gas integrals.
It is likely that the recipe for transforming the Eynard formalism 
into an effective  field theory proposed by Flume {\it et al.} \cite{Flume:2009aa} is the way to obtain this correspondence.

\bigskip
\leftline{\bf Acknowledgments}

\noindent  The author thanks A. Alexandrov, B. Eynard and  N. Orantin
for useful discussions.

  

   \footnotesize
   
   \providecommand{\href}[2]{#2}\begingroup\raggedright\endgroup

  \end{document}